\documentclass[twocolumn,showpacs,preprintnumbers,amsmath,amssymb]{revtex4}
\usepackage{graphicx}
\usepackage{amsmath}
\usepackage{dcolumn}% Align table columns on decimal point
\usepackage{bm}% bold math
\usepackage[dvips]{epsfig}
\begin{document}
\title{Three-dimensional MHD simulations of in-situ shock formation in the 
coronal streamer belt}
%%% ----------------------------------------------------------------------
\author{Yu. Zaliznyak\footnote{also at Institute for Nuclear Research, Kiev, Ukraine}}
 \email{Y.Zaliznyak@rijnh.nl}
\author{R. Keppens}%
\email{keppens@rijnh.nl}
\author{J.P. Goedbloed}
\email{goedbloed@rijnh.nl} 
\affiliation{FOM-Institute for Plasma Physics Rijnhuizen, 
              P.O.Box 1207, 3430 BE, Nieuwegein, The Netherlands}
\date{\today}
%-----------------------
\begin{abstract}
We present a numerical study of an idealized 
magnetohydrodynamic (MHD) configuration 
consisting of a planar wake flow embedded into a three-dimensional (3D) sheared 
magnetic field.
Our simulations investigate the possibility for in-situ development 
of large-scale compressive disturbances at cospatial current sheet -- velocity 
shear regions in the heliosphere. 
Using a linear MHD solver, we first systematically chart the
destabilized wavenumbers, 
corresponding growth rates, and physical parameter ranges 
for dominant 3D sinuous-type instabilities in an equilibrium
wake--current sheet system. 
Wakes bounded by sufficiently supersonic (Mach number $M_s > 2.6$) flow
streams are found to support dominant fully 3D sinuous instabilities 
when the plasma beta is of order unity.
Fully nonlinear, compressible 2.5D and 3D MHD simulations show the 
self-consistent formation of 
shock fronts of fast magnetosonic type.
They carry density perturbations far away from the wake's center. Shock 
formation conditions are identified in sonic and Alfv\'enic Mach number
parameter space. Depending on the wake velocity contrast and 
magnetic field magnitude, as well as on the initial perturbation, the 
emerging shock patterns can be plane-parallel as well as fully 
three-dimensionally structured. Similar large-scale transients could 
therefore originate
at distances far above coronal helmet streamers or at the location of 
the ecliptic current sheet.
\end{abstract}
%----------------------------
%----------------------------
\maketitle
%%% ----------------------------------------------------------------------
\section{Introduction}
\begin{figure}[ht]
\begin{center}
 {\resizebox{0.49\textwidth}{!}{\includegraphics{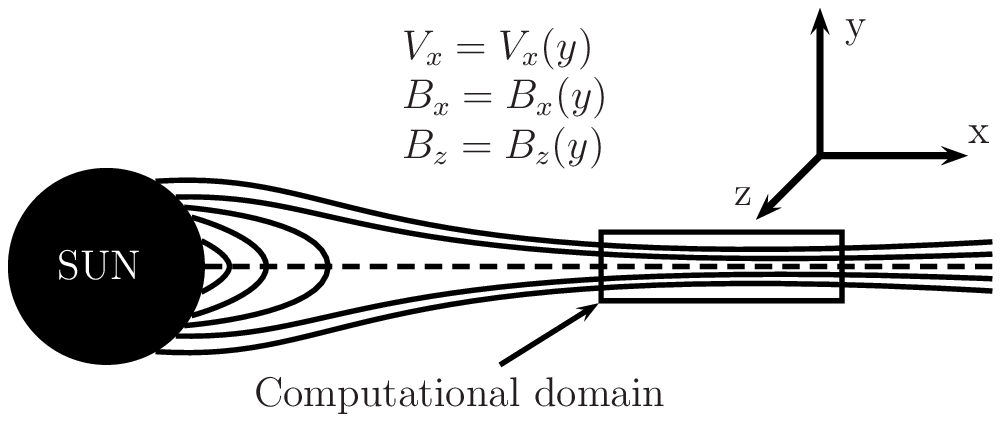}}}
  \caption{Model of the coronal streamer belt: a schematic view.}
\label{ModelFig}
\end{center}
\end{figure}

%%% ----------------------------------------------------------------------
\label{SectionIntroduction}
%%%%%%%%%%%%%%%%%%%%%%%%%%%%%%%%%%%%%%%%%%%%%%%%%%%%%%%%%%%%%%%%%%%%%
The magnetic and plasma flow configuration in a solar coronal streamer belt
beyond its helmet cusp is characterized by a region of cospatial
magnetic shear and a cross-stream velocity variation reminiscent of a `wake'
flow. Suitably simplified, a local study of streamer belt dynamics may
then consider a `boxed' model of a fluid wake -- current sheet configuration
as sketched in Fig.~\ref{ModelFig}. A large number of
previous studies have identified the various linear MHD instabilities
supported by such an equilibrium configuration, with purely
incompressible studies~\cite{EinaudiBoncinelliDahlburgKarpen1999}, 
as well as extensions to the compressible 
regime~\cite{DahlburgKeppensEinaudi2001,WangLeeWeiAkasofu1988,EinaudiChibbaroDahlburgVelli2001,DahlburgEinaudi1999}.
Their applications cover a range of physical problems, as they are used to 
model dynamics at the heliospheric 
current sheet~\cite{WangLeeWeiAkasofu1988}, mechanisms for variable slow solar
wind formation and acceleration~\cite{EinaudiBoncinelliDahlburgKarpen1999}, 
and plasmoid formation and evolution in the solar streamer 
belt~\cite{EinaudiChibbaroDahlburgVelli2001,DahlburgKarpen1995,Einaudi1999}.

Similarly idealized configurations
equally containing velocity shear layers co-spatial with an
electric current sheet or neutral sheet have been used to investigate
the linear and nonlinear stability properties of planar and non-planar
current -- vortex 
sheets~\cite{Dahlburg1998,ShenLiu1999,DahlburgBoncinelliEinaudi1997,AntognettiEinaudiDahlburg2002,DahlburgEinaudi2001,DahlburgEinaudi2000,OfmanChenMorrisonSteinolfson1991} 
and planar magnetized 
jets~\cite{Dahlburg1998,BiskampSchwarzZeiler1998,DahlburgKarpen1994,DahlburgBoncinelliEinaudi1998,OfmanChenMorrisonSteinolfson1991}. These configurations
were investigated in the context of solar flares and surges, among other
applications.
The possibility of having both shear-flow related instabilities, and
ideal and resistive magnetic instabilities, allow the plasma to transit
to turbulent regimes via various routes in its further nonlinear
evolution. Energy can be tapped from the background mean flow, 
with localized reconnection events releasing magnetic energy.

In this paper, we revisit the stability properties for a compressible
wake -- current sheet model of streamer belt dynamics, and extend earlier
two-dimensional nonlinear simulations~\cite{DahlburgKeppensEinaudi2001,EinaudiChibbaroDahlburgVelli2001,AntognettiEinaudiDahlburg2002}
to three-dimensional MHD computations. The necessity to model
the compressive aspects when applying the `box-model' from Fig.~\ref{ModelFig} 
to encompass both the slow and fast solar wind regions is obvious since
the wind becomes supersonic sufficiently close to the Sun (at roughly
$5\, R_\odot$, as follows from Parker's model 
\cite{Parker1958}). Compressibility can give rise to
new effects as compared to incompressible studies, 
such as nonlinear steepening of ideal sinuous instabilities under supersonic
flow conditions to form fast magnetosonic shock 
fronts~\cite{DahlburgKeppensEinaudi2001,AntognettiEinaudiDahlburg2002}, and the domination of 
fully three-dimensionally structured unstable modes over 
two-dimensional ones \cite{DahlburgEinaudi2000,DahlburgKeppensEinaudi2001}. In the incompressible case,
the validity of Squires theorem~\cite{DahlburgKarpen1995} 
precluded the latter possibility and allowed
one to concentrate on 2D dynamics. It is the main result of this paper to show 
that (1) strongly supersonic magnetized wakes, characterized
by a plasma beta of order unity, also have dominant 3D sinuous instabilities
which (2) can lead to fully 3D structured fast magnetosonic shock fronts
in their nonlinear evolution. 

Earlier, formation and amplification of fast magnetosonic shocks was 
shown for supersonic super-Alfv\'enic wakes by means of numerical 
simulations~\cite{DahlburgKeppensEinaudi2001}. 
These shocks were also observed in studies of 2D current-vortex 
sheets in the transonic regime \cite{AntognettiEinaudiDahlburg2002}. 
We supplement these earlier findings by a complete parametric study of initial 
state parameters which lead to shock formation. Our findings that dominant
3D sinuous instabilities are possible correct earlier claims to the
contrary~\cite{DahlburgKeppensEinaudi2001} where wavenumber space was not
sufficiently explored. At the same time this calls for 
a study of the nonlinear development of a sinuous instability 
in three dimensions, which we perform as well. 

It was pointed out by Wang et al.~\cite{WangLeeWeiAkasofu1988}, on the basis
of a linear stability analysis of a similar wake -- current sheet configuration,
that instabilities may develop in-situ at the heliospheric
current sheet at radial distances of order 0.5 -- 1.5 AU. They
emphasized the importance of streaming
sausage modes leading to traveling magnetic islands (clouds) and
plasmoids. Our results focus on the streaming kink mode, and our
nonlinear simulations demonstrate that it can lead to shock-dominated
large-scale transients. This re-emphasizes the possibility that not all
satellite solar wind observations of such events 
need to be directly mapped back to
a specific solar coronal event (eruptive prominence or CME). 

This paper is organized as follows. Section~\ref{SectionModelEQS} presents
the system of MHD equations describing the wake-current sheet evolution and 
the initial equilibrium configuration studied. 
Section~\ref{NumericalTools} 
describes the numerical tools used for the
linear stability analysis and for the nonlinear evolution studies.
In Section~\ref{SectionLinear}, we
present the linear stability analysis of the wake-current sheet
configuration and identify the previously unexplored
domain of the dominating three-dimensional sinuous instability. In 
Section~\ref{SectionNonlinear}, the nonlinear 
MHD simulations with the Versatile Advection Code~\cite{Toth1996} for
different values of wake velocity contrast and magnetic field magnitude
are discussed.
Formation of fast magnetosonic shocks is demonstrated and their properties and
generation conditions are described. 
Conclusions and a discussion of the relevance to
in-situ streamer belt dynamics are placed in Section~\ref{SectionConclusions}.

%%%%%%%%%%%%%%%%%%%%%%%%%%%%%%%%%%%%%%%%%%%%%%%%%%%%%%%%%%%%%%%%%%%%%
\section{MHD equations and equilibrium model}
\label{SectionModelEQS}
%%%%%%%%%%%%%%%%%%%%%%%%%%%%%%%%%%%%%%%%%%%%%%%%%%%%%%%%%%%%%%%%%%%%%

The system under consideration is described by the set of one-fluid
compressible resistive MHD equations of the form:
%%%%%
\begin{eqnarray}
 \nonumber
\partial_t \rho + \nabla \cdot \left( \rho {\bf v} \right) = 0, \\
 \nonumber
\partial_t \left( \rho {\bf v} \right) 
  + \nabla \cdot \left( {\bf v} \rho {\bf v} - {\bf B} {\bf B} \right) 
  + \nabla p_{tot} = 0, \\
 \nonumber
\partial_t e 
  +\nabla \cdot \left( {\bf v} e -{\bf BB}\cdot{\bf v} + {\bf v} p_{tot}\right)
  = \nabla \cdot \left( {\bf B} \times \eta {\bf J} \right),\\
 \nonumber
\partial_t {\bf B} + \nabla \cdot \left( {\bf vB-Bv} \right) 
  = -\nabla \times \left( \eta {\bf J} \right), \\
 \nonumber
{\bf J}=\nabla \times {\bf B}, \\
 \nonumber
p=\left(\gamma-1\right) \left(e - \rho v^2/2 -B^2 /2 \right), \,\, p_{tot} = p+B^2/2,
\end{eqnarray}
where $\rho$ is the mass density, ${\bf v}$ is the plasma velocity, 
${\bf B}$ is the magnetic field, and $e$ stands for total energy density. Other
quantities are ${\bf J}$ -- electric current, thermal pressure $p$ and the
constants $\gamma\equiv 5/3$ -- adiabatic index, and $\eta$ -- resistivity coefficient.
The resistive source terms are taken along to accurately simulate 
the small-scale reconnection events which only occur far into the 
nonlinear regime in the cases studied below. 

Guided by the box model of Fig.~\ref{ModelFig}, the magnetized wake flow
co-spatial with the current sheet can be modeled as a force-free equilibrium
configuration:
\begin{eqnarray}
V_x=1-{\rm cosh} ^{-1}y, \,\, V_y=0, \,\, V_z=0 \nonumber \\
B_x=M_a^{-1}{\rm tanh}\frac{y}{w}, \,\, B_y=0, \,\, B_z=M_a^{-1}{\rm cosh}^{-1}\frac{y}{w} 
\nonumber \\
\rho=1.0, \,\, p=(\gamma M_s^2)^{-1},
\label{WakeEQS}
\end{eqnarray}
where $M_s$ and $M_a$ are the sonic and Alfv\'enic Mach numbers for the
fast flow streams, respectively, and $w$
describes the thickness of the current sheet relative to the
width of the wake flow. The latter, together with the density and velocity
of the fast flow streams, has been used to define our unit system. 
The plasma $\beta$ is then
found from $\beta=(2M_a^2)/(\gamma M_s^2)$. If not mentioned otherwise,
we set $w=1$. The same force-free equilibrium was 
analyzed in~\cite{DahlburgKeppensEinaudi2001}, while a similar, but non 
force-free magnetic configuration,
was used in~\cite{WangLeeWeiAkasofu1988}. 

%
%%%%%%%%%%%%%%%%%%%%%%%%%%%%%%%%%%%%%%%%%%%%%%%%%%%%%%%%%%%%%%%%%%%%%
\section{Numerical tools}
\label{NumericalTools}
%%%%%%%%%%%%%%%%%%%%%%%%%%%%%%%%%%%%%%%%%%%%%%%%%%%%%%%%%%%%%%%%%%%%%
Linear stability analysis of the wake-current sheet configuration was 
performed numerically with the LEDAFLOW 
code~\cite{NijboerHolstPoedtsGoedbloed1997}. This code can
compute the complete MHD spectrum of all waves and instabilities that are
eigenfrequencies of one-dimensionally varying 
equilibria containing background plasma flows. The linearized MHD equations are
discretized using finite elements (quadratic and cubic Hermite polynomials) 
in the
direction of inhomogeneity and a Fourier representation of the
perturbations in the invariant directions.
Assuming a temporal variation as $\exp(\lambda \, t)$, application of the
Galerkin procedure results 
in a complex non-Hermitian eigenvalue problem which is then 
solved by a complex QR-decomposition-based solver. For each isolated 
eigenvalue $\lambda$, the set of corresponding eigenfunctions can be found 
with an inverse vector iteration. 
In the QR mode, the whole spectrum for a given set of Fourier mode numbers
is calculated, but the amount of grid points in the finite element
representation is usually limited to 100 -- 120 points. In 
inverse iteration mode, the spatial resolution is much higher to be able to 
get converged eigenfrequencies with strong gradients in their eigenfunctions
(up to 5000 points). The boundary conditions implemented in LEDAFLOW assume 
perfectly conducting walls at $y$-boundaries, which is, strictly speaking, not
applicable to the streamer belt configuration. 
To prevent any influence of these conducting
walls on the solution of the eigenvalue problem, one should place them at
sufficiently large distances in the
perpendicular $y$-direction, and check whether the eigenfunctions are 
localized away from the
computational domain boundaries and vanish close to the walls.

Nonlinear modeling of the wake-current sheet dynamics was performed by
the general finite-volume-based Versatile Advection Code 
(VAC) \cite{Toth1996, Keppens2001}. For all runs we used the full set of 
compressible resistive 
MHD equations with a constant resistivity $\eta=10^{-4}$ value. 
The one-step total
variation diminishing (TVD) method \cite{Harten1983JComputPhys} was employed, 
which is a second-order accurate shock-capturing scheme. It uses an
approximate Roe-type \cite{Roe1981} 
Riemann solver with the minmod (for 3D simulations) or Woodward (for 2.5D
simulations) limiting applied to the characteristic waves. 

The rectangular computational domain has periodic boundaries in 
the streamwise $x$-direction, and open
boundaries in the cross-stream direction $y$, such that perturbations
reaching these boundaries are allowed to propagate freely outside the
computational box. 
For 3D runs, the spanwise direction $z$ is periodic as well.
We used non-equidistant rectangular grids with a symmetric 
(around $y=0$) accumulation near the core of the wake. Grid
resolutions varied from $100 \times 200$ up to $300 \times 600$ in 2.5D and 
from $60 \times 90 \times 60$ up to $100 \times 150 \times 100$ in 3D cases. 
The grid lengths in the streamwise and spanwise directions are found from
$L_x=2\pi/k_x$ and $L_z=2\pi/k_z$, respectively, with wavenumbers
$k_x, \,k_z$ corresponding to the most unstable mode 
identified from the linear stability analysis. For the
cross-stream boundary, the open boundaries are placed at
$L_y=\pm 25$--$50$ depending on the localization 
of the unstable eigenmode. 2.5D runs were performed 
on an SGI Octane workstation, 
while 3D simulations were executed on typically 32 processors
of the SGI Origin 3800 at the Dutch Supercomputing Centre, using openMP for
parallelization purposes.
%
%%%%%%%%%%%%%%%%%%%%%%%%%%%%%%%%%%%%%%%%%%%%%%%%%%%%%%%%%%%%%%%%%%%%%
\section{Linear stability analysis}
\label{SectionLinear}
%%%%%%%%%%%%%%%%%%%%%%%%%%%%%%%%%%%%%%%%%%%%%%%%%%%%%%%%%%%%%%%%%%%%%
Depending on the parameters of the one-dimensionally varying (but fully
three-dimensionally structured) equilibrium configuration given 
by Eqs.~(\ref{WakeEQS}), as well
as on the value of the resistivity and the lenghtscales of the imposed 
perturbation, up to three different mode types
may be destabilized in the 
wake-current sheet configuration. These are known as
the ideal sinuous (or streaming kink) mode, and an 
ideal and a resistive varicose (sausage-type) 
mode~\cite{Einaudi1999,WangLeeWeiAkasofu1988}. 
In a 2D case these types are easily
distinguishable by the symmetry in their cross-stream velocity perturbations, 
which is even for the sinuous mode and odd for both varicose modes. 
In a 3D case these symmetry properties are broken. 
Here we shall mainly focus at the instability of an 
ideal sinuous type, already present in unmagnetized hydrodynamic wake
configurations. 
In the range of Mach and Alfv\'en Mach numbers investigated, its growth rate is 
always larger than 
that of the varicose modes, and thus the sinuous mode is expected to 
dominate the evolution of the system. This mode is also known to
demonstrate interesting nonlinear evolution: in 2.5D MHD simulations,
sinuous mode instabilities led to the self-consistent formation of fast
magnetosonic shocks in the nonlinear stage~\cite{DahlburgKeppensEinaudi2001}.

\begin{figure}[ht]
\begin{center}
 {\resizebox{0.49\textwidth}{!}{\includegraphics{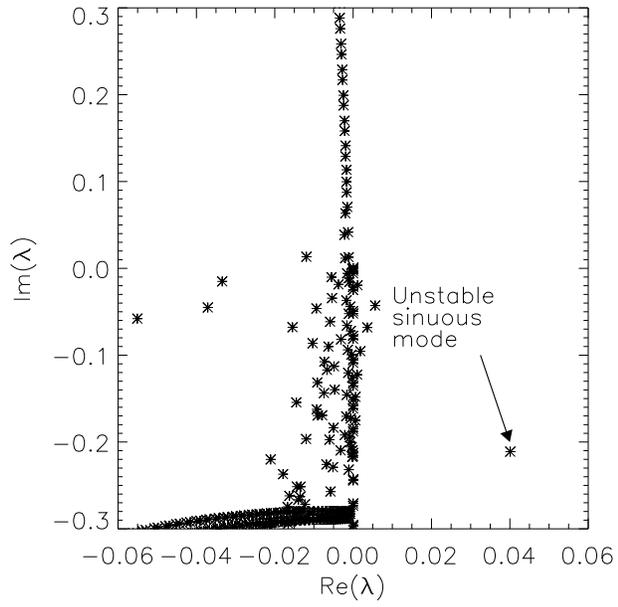}}}
  \caption{MHD spectrum of a compressible wake-current sheet with 
           $M_s=3$, $M_a=5$, $k_x=0.35$, $k_z=0.15$, $\eta=0.01$. 
	   The unstable sinuous mode is marked.}
\label{LinearSpectra}
\end{center}
\end{figure}

In Fig.~\ref{LinearSpectra} a typical spectrum of a wake-current sheet
system obtained with LEDAFLOW is presented. The
parameters used are for a supersonic $M_s=3$, super-Alfv\'enic $M_a=5$ wake at
fixed streamwise wavenumber $k_x=0.35$ and spanwise wavenumber $k_z=0.15$.
The figure shows individual
complex eigenfrequencies $\lambda$ throughout the complex plane,
but for the value of the resistivity $\eta=0.01$, the ideal and 
resistive varicose instabilities are fully suppressed, and only the
ideal sinuous mode remains unstable in the system (marked by the arrow
in Fig.~\ref{LinearSpectra}). 
Since the sinuous instability is ideal in nature, 
its growth rate does not depend on the value of the resistivity chosen.
The eigenfunctions, namely the perturbed
plasma density $\delta \rho$ and cross-stream velocity $\delta v_y$, for the
unstable sinuous mode are shown in 
Fig.~\ref{EigenFunctions}. 
The instability growth rate is $\gamma \approx 0.0402$ and the phase velocity 
$v_{ph} \approx 0.55$. It is seen that 
$\delta v_y$ is nearly symmetric (even) with respect to $y=0$, while the 
density perturbation has a more complicated structure but is in essence odd.
All linear stability results following are performed for a fixed value of 
resistivity as used in the nonlinear simulations, namely $\eta=10^{-4}$.

The maximal sinuous mode growth rate is achieved for an incompressible
(limit $M_s\rightarrow 0$), purely hydrodynamical (\textbf{B}=0) case. In
incompressible magnetized wakes, the dominant instability
was found to be two-dimensional having zero spanwise wavenumber, 
in agreement with the Squires theorem~\cite{DahlburgKarpen1995}.  
When plasma compressibility is taken into account, as done in this paper,
the situation can change drastically. In~\cite{DahlburgKeppensEinaudi2001},
it was demonstrated that at high Mach numbers, an increase of the
spanwise wavenumber can make the ideal compressible varicose mode more unstable.
The sinuous mode growth rate, however, was believed to decrease with increasing
$k_z$.

In Fig.~\ref{LinGrowthRates}, the numerically obtained growth rates 
for $M_s=3$, $M_a=5$ wakes of
the ideal sinuous instability are plotted versus streamwise wavenumber $k_x$ 
(left panel A), and spanwise wavenumber $k_z$ (right panel B). 
The top curve in panel~(A) for $k_z=0$ is identical to 
the corresponding dispersion
curve for Mach~3 wakes shown in~\cite{DahlburgKeppensEinaudi2001}, their 
Fig.~1(b). It is clearly seen from panel~(A) that the
region of destabilized $k_x$ is widest in the 2D case ($k_z=0$). 
Increasing the spanwise wavenumber $k_z$ leads to a shrinking of the
instability domain from both sides, while for increasing $k_z$, 
the most unstable streamwise wavenumber also increases.
This trend continues up to some upper value of $k_z > 0.6$ where the
sinuous mode becomes stable. Panel~(B) demonstrates that the 2D landscape
of growth rate versus streamwise $k_x$ and spanwise $k_z$ wavenumber,
has a secondary `ridge' which was overlooked 
in~\cite{DahlburgKeppensEinaudi2001}. There, at a fixed wavenumber $k_x=0.55$,
an increase of the spanwise wavenumber led to a decreasing growth
rate for the sinuous mode (their Fig.~3(b)). Our panel~(B) shows that
their study overlooked the rather interesting case of larger streamwise
wavelength (smaller $k_x$), where three-dimensional local 
extremum of growth rate was found for $k_x < 0.25$. We can further observe that with an 
increase of the streamwise wavenumber $k_x$ from 0.075 up to values 0.35, 
both the growth rate and the instability existence domain (in $k_z$) increase, 
reaching their maximum values at $k_x\approx 0.35$, which corresponds
to maximal 2D growth. Further increasing $k_x$
leads to a renewed drop in growth rate, however the instability domain does not
change significantly. The latter can be seen when comparing our $k_x=0.35$
curve from panel~(B) with the $k_x=0.55$ curve for Mach~3 wake flow 
in~\cite{DahlburgKeppensEinaudi2001}, their Fig.~3(b).
Below some $k_x < 0.075$ the sinuous mode is fully stabilized again.

The importance of this secondary ridge in the growth rate landscape is
marginal for the $M_s=3$, $M_a=5$ flow conditions of 
Fig.~\ref{LinGrowthRates}. Indeed, the absolute maximal growth rate corresponds
to a purely 2D ($k_z=0$) mode with $k_x\approx 0.35$. 
However, for different combinations of
Mach and Alfv\'en Mach number, the most unstable mode can be fully 3D in nature,
a fact completely missed by earlier studies of this wake system. 
Variations of ideal sinuous mode
maximal growth rate versus spanwise wavenumber for different values of $M_s$
and $M_a$ are plotted in Fig.~\ref{GammaMax_Ms_Ma}, and the possibility of
a 3D dominant instability is clearly manifested. Note that in this figure,
the streamwise wavenumber $k_x$ varies from curve to curve, 
and has a value corresponding to the overall maximum 
in the growth rate landscape.

For a given Mach
number, increasing the magnetic field strength (decrease of Alfv\'en Mach
number $M_a$) acts to suppress the sinuous instability. 
This is also seen in Fig.~\ref{GammaMax_Ms_Ma}, panel~(A).
The range of
destabilized $k_x$ decreases, and the streamwise
wavelength of the fastest growing mode decreases. 
When $M_a$ becomes smaller than some critical value $M_a^{cr}$, the system
is stable.
This critical value $M_a^{cr}$ was found to be approximately equal to
2.5, and this value is rather universal in a wide range of studied cases
($0<M_s<10$). This is consistent with~\cite{DahlburgKeppensEinaudi2001}, 
their Fig.~2. Note that Wang et al.~\cite{WangLeeWeiAkasofu1988} found a lower
cut-off value (namely $M_a^{cr}\approx 1.2$) which must be due to the
different equilibrium configuration used. Their equilibrium
magnetic field strength is weaker at the center of the flow wake, while our
configuration has a constant field magnitude throughout. Therefore, the
stabilizing influence of magnetic tension translates into a lower
cut-off value for $M_a^{cr}$.

Generally, with an increase of the sonic Mach number $M_s$, the overall
maximal growth rate of the sinuous instability goes down, 
as well as the diapason of destabilized 
wavenumbers. For an Alfv\'en Mach number $M_a=3.5$, variations from Mach 2
till Mach 15 flows on the sinuous growth rates
are shown in panel~(B) from Fig.~\ref{GammaMax_Ms_Ma}.
Corresponding dependences ${\cal R}{\rm e} \left[\lambda (k_x)\right]$ 
can be found elsewhere, e.g. see Fig.~1(b) in~\cite{DahlburgKeppensEinaudi2001}.

The most important finding from Fig.~\ref{GammaMax_Ms_Ma} is that 
the most unstable sinuous mode 
may become oblique to the shear layer in a wake -- current sheet system 
for sufficiently supersonic flows. Untill now, this was
known for the compressible 
plane current-vortex sheet \cite{DahlburgEinaudi2000} as well as for
varicose-type modes in compressible wake -- current sheet configurations 
\cite{DahlburgKeppensEinaudi2001}. 
We found that for a
given sonic Mach number larger than $M_s \approx 2.6$, there exists a range
in $M_a$ where the maximum growth occurs for a mode with nonzero spanwise
wavenumber. That diapason of $M_a$ for dominant
three-dimensional instability
is bounded from below and from above, and most wide near $M_s \approx 2.6$ where
it is $2.9<M_a<4$. Its size decreases slowly 
with increasing $M_s$, so that even for highly supersonic wake flows
the 3D instability domain is non-negligible,
e.g. for $M_s=10$ it is found in the range $2.65<M_a<3.025$. 

In the next sections we will study the nonlinear evolution of the
wake--current
sheet system in 2D and 3D using the information about the most unstable
modes as calculated by LEDAFLOW for constructing
the initial conditions.

%%%%%%%%%%%%%%%%%%%%%%%%%%%%%%%%%%%%%%%%%%%%%%%%%%%%%%%%%%%%%%%%%%%%%
\section{Nonlinear evolution and shock formation}
\label{SectionNonlinear}
%%%%%%%%%%%%%%%%%%%%%%%%%%%%%%%%%%%%%%%%%%%%%%%%%%%%%%%%%%%%%%%%%%%%%
  \subsection{2.5D MHD parameter study of shock formation}
%%%%%%%%%%%%%%%%%%%%%%%%%%%%%%%%%%%%%%%%%%%%%%%%%%%%%%%%%%%%%%%%%%%%%
Since we are mainly interested in the nonlinear evolution of sinuous-type 
modes which have the largest growth rates and therefore dominate the 
system, the initial equilibrium configuration given by Eqs.~(\ref{WakeEQS}) 
was perturbed
by a symmetric cross-stream velocity perturbation of the form:
\begin{displaymath}
\delta v_y \sim \varepsilon \sin \left( k_x x \right) \exp(-y^2),
\end{displaymath}
with typical values of $\varepsilon = 10^{-3}..10^{-2}$.
A comparative analysis of the nonlinear wake response 
to such perturbations in 2.5D MHD,
for varying sonic and Alfv\'en Mach numbers, was presented 
in~\cite{DahlburgKeppensEinaudi2001}. Here, we will focus on previously
unaddressed issues, namely the interplay between simultaneously 
excited modes having different lenghtscales and growth rates, 
and the properties and formation conditions for shock waves, 
which are observed in 
the simulations at the instability saturation stage.

\subsubsection*{Mode-mode interplay}

In Fig.~\ref{Plate25D}, an example of a 2.5D MHD simulation is presented.
By definition, all 2.5D simulations include only $k_z=0$ modes.
Initially, two sinuous modes with streamwise wavenumbers
$k_{x1}=0.35$ (mode 1) and 
$k_{x2}=0.35/3=0.11667$ (mode 2) were
excited on the grid with amplitudes $\varepsilon_1 = 2.5\cdot 10^{-5}$ and 
$\varepsilon_2= 10^{-2}$, respectively. 
For the simulated $M_s=3$, $M_a=5$ ($\beta=3.33$) wake, 
mode~1 corresponds to the dominant
2D fastest growing mode in the system. Note that we perturb the slower growing
mode at a higher amplitude to force mode interaction. Fig.~\ref{Plate25D}
shows snapshots of density and cross-stream velocity, while for the
same time frames, Fig.~\ref{Spectra25D} presents the spatial spectrum 
of the cross-stream velocity.
This spatial Fourier transform performed for the sequence of snapshots 
allows to extract information on the linear growth rate for each mode
and to confirm LEDAFLOW predictions. The linear results of the
previous section were
found to be very accurate, the difference in growth rates deduced 
from VAC simulations and those obtained by LEDAFLOW is usually less than 1\%.

At early times, mode~2 is seen to grow with the growth rate $\approx 0.018$. 
However, at times around $T=200$, as it is clearly seen in the spectral evolution, 
mode~1 obtains a comparable amplitude. Having the
larger growth rate, mode~1 develops faster, and at times around $T=250$ its
amplitude becomes larger than that of mode~2. Later on in the nonlinear
evolution the short-wavelength mode~1 remains the most dominant one, but a 
cascade to smaller wavelength dynamics is obviously occuring. Note that 
initially
(times $T<200$) this only involves direct overtones of the two excited modes.

Similar observations can be made as follows. In the linear stage of the
instability, the volume average cross-stream 
kinetic energy in the system plotted in Fig.~\ref{Energies25D}, and defined as 
\begin{equation}
E^y_{Kin}=\left( 2L_xL_yL_z \right)^{-1} 
     \int_{0}^{L_x} \int_{-0.5L_y}^{0.5L_y} \int_{0}^{L_z} \rho v_y^2 dx dy dz,
\end{equation}
grows exponentially in accordance with the slower growing mode~2. Again, its
slope changes near $T=200$ indicating that the faster growing mode~1 comes into 
play. Eventually, transverse kinetic energy growth is halted and
this saturation point typically coincides with the time where the
largest density contrast is achieved.
The total kinetic energy of the system decreases, while the magnetic energy
increases indicative of the occuring energy conversion processes. 

\subsubsection*{Tresholds for shock formation}

Depending on the system parameters, exactly at the point of
maximal density contrast, shock waves may be formed. They are
clearly detected in the density snapshot at time $T=300$ from 
Fig.~\ref{Plate25D}. Note that these shocks extend far out into the fast
stream regions, up to 25 wake widths in the cross-stream direction.
Fig.~\ref{RHOCut25D} shows a cross-cut of the density variation at this time,
taken at about
10 wake widths above the wake-current sheet. The shocks carry strong
density contrasts, up to a factor of 2. Note that due to the mode-mode
interaction, successive shock fronts can have different strengths, and
their interseparation no longer corresponds to the wavelength of the
linearly most unstable mode. This cross-cut can loosely be interpreted
as a time-evolution of an in-situ satellite measurement, since the shock
fronts are typically advected over the periodic side boundaries for more
than a full advection cycle.

To identify the type
of the shock front, we used Rankine-Hugoniot relations to determine the
shock speed. This is most easily found from mass continuity, where 
the velocity of the shocks is found as 
$V_{sh}=(\left(\rho v_{\perp}\right)^{before}-\left(\rho v_{\perp}\right)^{after})/(\rho^{before}-\rho^{after})$. 
The shock type then follows from determining 
Alfv\'en, slow- and fast magnetosonic Mach number transitions 
in the coordinate system which moves with the shock velocity.
Shock waves formed at the nonlinear stage of the sinuous instability
of a supersonic magnetized wake were found to be fast magnetosonic in nature.

The steepening of the sinuous wave fronts into shock waves is possible 
only above a threshold value of sonic Mach number. That threshold depends on the amplitude
of the magnetic field (hence on $M_a$). In a pure hydrodynamical case
($M_a=\infty$), sonic shocks are formed for supersonic wake 
flows with $M_s \ge 2.10$. Shocks appeared for all studied hydrodynamic
cases in the range $2.10 < M_s < 10.0$ and no upper limit of $M_s$ for
their formation was found. This hydro critical value is recovered in
weak ($M_a>5$) magnetized wakes. However, with an increase of 
the magnetic field magnitude, the minimum value $M_s^{Min}$ of the
sonic Mach number needed for shock formation, is increased. This is
presented in Fig.~\ref{ThresholdMs}, where threshold sonic Mach numbers 
are plotted versus the
Alfv\'en Mach numbers. Formation of the shock fronts was shown to be possible
for any $M_a$ where the sinuous mode is still unstable ($M_a>2.5$).
Hence, for any magnetic field strength, where the sinuous instability is not 
suppressed linearly, we were able to identify a shock-treshold value of 
Mach number $M_s^{Min}$. For all $M_s>M_s^{Min}(M_a)$ shock fronts form
self-consistently through wave front steepening.

Most of the nonlinear runs to determine the parameter ranges for shock
formation were performed for systems of length $L_x=2\pi/k_x^m$, with
$k_x^m$ being the streamwise wavenumber of the fastest growing mode. 
It turns out that shocks can appear having different scales in the
x-direction, since there
exists a diapason of streamwise wavenumbers $k_x$, located around $k_x^m$, for
which shocks develop. At scales sufficiently far from $k_x^m$,
formation of shock waves was not observed. For a given Alfv\'en Mach 
number, with an increase of the wake speed contrast
($M_s$), the angle between the shock front and the streamwise 
direction decreases. Hence the density perturbations become more 
and more cuddled up to the core of the wake, and thus not always reach
far cross-stream regions.

Finally, when applied to the coronal streamer
belt, the width of the current sheet -- $w$ in Eqs.~(\ref{WakeEQS}) --
is typically much smaller
than that of the fluid wake~\cite{EinaudiChibbaroDahlburgVelli2001}. 
With a decrease of the relative width of the current sheet, the number of grid points
needed to achieve acceptable resolutions grows rapidly, and simulations become
more and more time- and memory consuming. Nevertheless, for selected cases we did
vary the current sheet width, and these experiments did not show any 
qualitative difference in the nonlinear stage of evolution, indicating that 
the model of equal-sized wake and current sheet used throughout
this paper is warranted. Quantitatively, as can also be obtained from a
linear LEDAFLOW analysis, the growth rate and streamwise wavenumber of 
the most unstable sinuous mode decreases when the current sheet width goes down.

%%%%%%%%%%%%%%%%%%%%%%%%%%%%%%%%%%%%%%%%%%%%%%%%%%%%%%%%%%%%%%%%%%%%%  
\subsection{Case studies in 3D MHD}
%%%%%%%%%%%%%%%%%%%%%%%%%%%%%%%%%%%%%%%%%%%%%%%%%%%%%%%%%%%%%%%%%%%%%  

The 2.5D simulations in the previous section excluded the emergence of
fully 3D structured shock fronts. However, our linear stability study demonstrated
the existence of dominant 3D unstable modes. Here we discuss 
three particular three-dimensional runs 
performed for different parameters of the equilibrium configuration and for
different initial excitations. We investigate two distinct cases:
(1) a wake with a dominant purely 2D mode, where we look into the
possibility of 3D shock structuring due to mode-mode interactions; and (2)
a wake previously identified to have a dominant 3D linear mode.

\subsubsection*{Mode-mode interaction}

In Fig.~\ref{Plate3D}, plasma density isosurfaces are plotted for different
stages in the sinuous mode instability development. In the first column
of Fig.~\ref{Plate3D}, values of
sonic and Alfv\'en Mach numbers were chosen to be $M_s=3$, $M_a=5$, as in the
2.5D run from Fig.~\ref{Plate25D}. The wake 
was perturbed with two modes: a pure 2D mode -- the most
dominant instability -- having $k_x=0.35, \,\, k_z=0$, and a
3D mode $k_x=0.35/3=0.11667, \,\, k_z=0.125$. The amplitude
of the 3D mode ($10^{-2}$) was much larger than that of the 2D mode ($10^{-5}$).
At first, the 3D sinuous instability develops in accordance with its
growth rate $\approx 0.018$, and the emerging density structure is clearly
three-dimensional. However at times around $T \approx 175$ (middle snaphot),
the 2D mode comes into play. It leads to elongation of the density
structures along
the $z$-axis. It again causes a slope change in the $E^y_{kin}(t)$ dependence 
plotted in Fig.~\ref{Energies3D}.
This continues untill the wake density structure tends to be mainly
two-dimensional, with very slight modulations along the $z$-axis. In particular,
the emerging shocks are nearly plane-parallel, as in the 2.5D
simulations. 

If the same system is 
initially perturbed by the 3D mode alone, as is done in the middle column 
in Fig.~\ref{Plate3D}, one can obtain fully 3D structured shock-dominated
wake dynamics. The two-dimensional mode starts to grow from the noise level
in this case, and obtains noticeable amplitude only at times 
when the 3D instability is close to saturation. The
shock fronts have a truly 3D topology. Note that at times 
around $T=250$ each high density blob 
splits into parts whose scale corresponds to the scale of the
fastly growing 2D $k_x=0.35, \,\,k_z=0$ mode. After saturation, a set of
nearly two-dimensional (as seen from the third snapshot in the middle
column in Fig.~\ref{Plate3D}) structures form. The number of structures
corresponds to the scale of the initially excited 3D mode, but their size in 
the streamwise direction is strongly influenced by the 
$k_x=0.35, \,\,k_z=0$ mode. These two runs demonstrate that 3D shock
structuring can even emerge from wakes with dominant 2D linear instabilities,
but only under rather selective initial excitations.

\subsubsection*{Dominant 3D instability}

The last column in Fig. \ref{Plate3D} shows the nonlinear
evolution of the density in pure three-dimensional
dynamics of the sinuous instability for the case $M_s=3$, $M_a=3.5$ ($\beta=1.6$). 
Above, it was demonstrated that for these parameters, 
the most unstable sinuous mode is oblique to the wake's 
core and has the wavenumbers $k_x=0.375, \,\, k_z=0.15$ 
(see Fig.~\ref{GammaMax_Ms_Ma}). Note how at all stages in the evolution,
the spatial density distribution remains fully 3D.
As compared to the other cases shown, the maximal
density contrast reached in the nonlinear stages is larger. 
For these parameters, the fast magnetosonic shock 
fronts emerging in the nonlinear stage have a fairly complex 3D structure.
The transverse kinetic energy evolution is plotted in Fig.~\ref{Energies3D}.

%%%%%%%%%%%%%%%%%%%%%%%%%%%%%%%%%%%%%%%%%%%%%%%%%%%%%%%%%%%%%%%%%%%%%
\section{Conclusions}
\label{SectionConclusions}
%%%%%%%%%%%%%%%%%%%%%%%%%%%%%%%%%%%%%%%%%%%%%%%%%%%%%%%%%%%%%%%%%%%%%
In this paper we performed a numerical study of the linear properties
and nonlinear evolution of
wake -- current sheet configurations by means of compressible resistive MHD
simulations. We focused on three-dimensional 
effects for the ideal sinuous-type instability. In contrast to previous 
studies, we have shown that 
there exist ranges of Mach number $M_s$ and Alfv\'en Mach number $M_a$, 
where the most unstable sinuous mode is oblique to the shear 
layer. Such pure three-dimensional instabilities develop nonlinearly 
into fully 3D deformations of the wake system containing fast
magnetosonic shock fronts.
For dominant 3D instabilities, the Mach number should exceed a
threshold value $M_s>2.6$. 
With increase of $M_s$, the range 
of $M_a$ corresponding to 3D instability decreases, but even 
for highly supersonic cases, a dominant 3D instability still exists in some 
range of $M_a$. At the nonlinear stage of this sinuous instability in 
the supersonic regime, 
self-consistent formation of fast magnetosonic shocks is observed.
They carry 
density jumps (by factors of 2) in cross-stream fast flow regions
far away from the wake center. As a result of mode-mode interaction,
the shock strengths and interseparations can vary.
Depending on the wake velocity contrast and magnetic field 
magnitude, as well as on the streamwise and spanwise scale of the initial 
perturbation, these shocks may have plane-parallel as well as complicated
three-dimensional structure. Shock fronts appear only for 
Mach numbers above some critical value, which depends on magnetic field 
strength, but is always larger than the hydro limit $M_s=2.1$. 

Finally, let us estimate the heliospheric regions, where our box-model 
predicts the possibility for in-situ compressible transients. 
Using the Parker model~\cite{Parker1958} for 
the interplanetary magnetic field one finds, in a manner originally  
done by Wang et al.~\cite{WangLeeWeiAkasofu1988},
Alfv\'en and sonic Mach number radial dependencies of the form: 
\begin{displaymath}
  M_s(r)=\frac{\delta v(r)}{v_a^0}
         \sqrt{\frac{2}{\gamma \beta_0}}\sqrt[4]{\frac{r^2}{r^2+1}},
\end{displaymath}
%%%%%
\begin{displaymath}
  M_a(r)=\frac{\delta v(r)}{v_a^0}\sqrt{\frac{2r^2}{r^2+1}},
\end{displaymath}
where $\delta v(r)$ is the 
velocity contrast between fast and slow wind streams, 
and $\beta_0$ and $v_a^0$ indicate the plasma beta and Alfv\'en speed at 1 AU.
For solar wind parameters at 
the Earth's orbit given by 
$n=5$ cm$^{-3}$, $B=5$nT, $T_e=2 \cdot 10^5$K, 
$T_p=4 \cdot 10^4$K (taken from~\cite{Winterhalter1994}), one finds 
$\beta_0 \approx 1.6$, while $v_a^0 \approx 50$ km/sec.
Assuming that the fast-slow solar wind velocity contrast,
as determined by Ulysses measurements~\cite{McComas1998} to be of the order
$\delta v \approx 350$km/sec, 
does not change significantly with radius beyond the various magnetosonic
transitions, one can compute $M_s(r)$ and $M_a(r)$ dependencies.

Our linear theory predicted that the sinuous mode is destabilized for 
$M_a \geq 2.5$ which then corresponds to
distances beyond 0.3 AU from the Sun. 
Relevant heliospheric distances turn out to be 0.3 AU -- 1.5 AU. 
Since we used the fluid wake scale length and the
fast-slow velocity contrast to define dimensionless variables, we can
verify whether the instability grows sufficiently fast to influence
conditions at 1 AU. The flow wake width is observationally
determined to be of order 300.000km at 1 AU, while
at a heliospheric distance of order 0.5 AU it is
estimated to be around 150.000km~\cite{Winterhalter1994}. 
Our time unit will then correspond to approximately 7 minutes. 
At 0.5 AU, the most unstable 
sinuous mode has a dimensionless growth rate of 0.0272 and corresponding
streamwise wavenumber of $k_x=0.325$. Hence,
the sinuous instability could fully develop within about 25 hours, in close
correspondence with the transit time to 1 AU. The lengthscale of these
compressive perturbations are then of order $2 \cdot 10^{-2}$AU. This
points to the distinct possibility of in-situ
shock formation at wake-current sheet configurations within the heliosphere. 

\vspace*{1cm}
%%%%%%%%%%%%%%%%%%%%%%%%%%%%%%%%%%%%%%%%%%%%%%%%%%%%%%%%%%%%%%%%%%%%%
{\bf Acknowledgements.}
%%%%%%%%%%%%%%%%%%%%%%%%%%%%%%%%%%%%%%%%%%%%%%%%%%%%%%%%%%%%%%%%%%%%%
This work was performed as part of the Computational Science programme 
$\,$ "Rapid Changes in Complex Flows" coordinated by JPG, and funded by the 
Netherlands Organization for Scientific Research (NWO). Computing facilities
were provided by NCF (Stichting Nationale Computerfaciliteiten). RK and
JPG performed this work supported by the European Communities under the 
contract of Association between EURATOM/FOM, carried out within the framework of
the European Fusion Programme. Views and opinions expressed herein do not
necessarily reflect those of the European Commission.
\bibliographystyle{unsrt}
\bibliography{ZaliznyakKeppensGoedbloed}
%
%%%%%%%%%%%%%%%%%%%%%%%%%%%%%%%%%%%%%%%%%%%%%%%%%%%%%%%%%%%%%%%%%%%%%
%%%%%%%%%%%%%%%%%%%%%%%%%%%%%%%%%%%%%%%%%%%%%%%%%%%%%%%%%%%%%%%%%%%%%
%%%%%%%%%%%%%%%%%%%%%%%%%%%%%%%%%%%%%%%%%%%%%%%%%%%%%%%%%%%%%%%%%%%%%
%%%%%%%%%%%%%%%%%%%%%%%%%%FIGURES%%%%%%%%%%%%%%%%%%%%%%%%%%%%%%%%%%%%
%%%%%%%%%%%%%%%%%%%%%%%%%%%%%%%%%%%%%%%%%%%%%%%%%%%%%%%%%%%%%%%%%%%%%
%%%%%%%%%%%%%%%%%%%%%%%%%%%%%%%%%%%%%%%%%%%%%%%%%%%%%%%%%%%%%%%%%%%%%
%%%%%%%%%%%%%%%%%%%%%%%%%%%%%%%%%%%%%%%%%%%%%%%%%%%%%%%%%%%%%%%%%%%%%
%
\clearpage
%%%%%%%%%%%%%%%%%%%%%%%%%%%%%%%%%%%%%%%%%%%%%%%%%%%%%%%%%%%%%%%%%%%%%
%%% \clearpage
%%%%%%%%%%%%%%%%%%%%%%%%%%%%%%%%%%%%%%%%%%%%%%%%%%%%%%%%%%%%%%%%%%%%%
\clearpage
\begin{figure}[ht]
\begin{center}
% {\resizebox{0.9\textwidth}{!}{\includegraphics{Fig3.eps}}}
  \caption{Eigenfunctions corresponding to the unstable sinuous mode from 
  Fig.~\ref{LinearSpectra}. Real and imaginary parts of the cross-stream 
  perturbed velocity and perturbed density are shown.}
\label{EigenFunctions}
\end{center}
\end{figure}
%%%%%%%%%%%%%%%%%%%%%%%%%%%%%%%%%%%%%%%%%%%%%%%%%%%%%%%%%%%%%%%%%%%%%
\clearpage
\begin{figure}[ht]
 {\resizebox{0.45\textwidth}{!}{\includegraphics{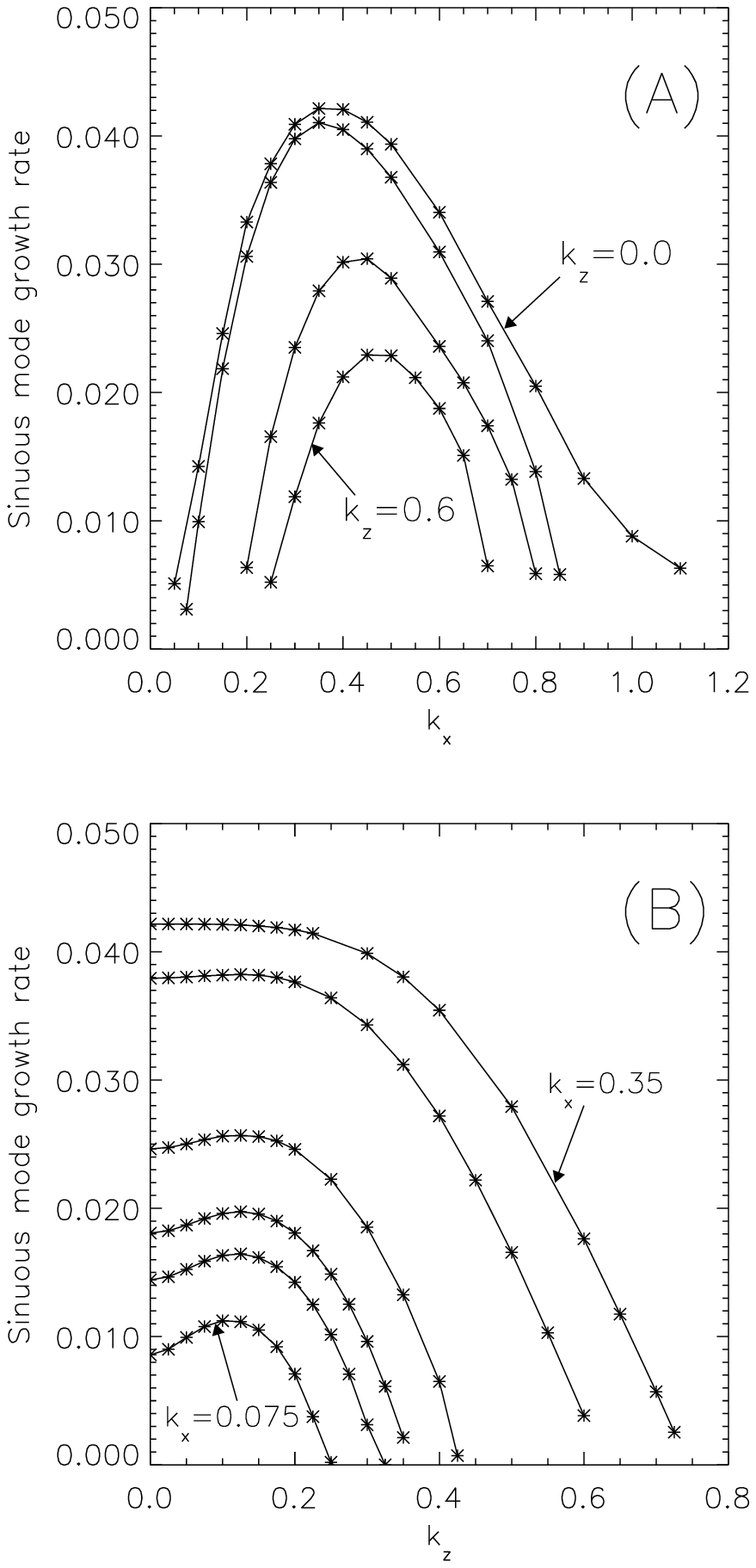}}}
  \caption{Ideal sinuous mode growth rate versus streamwise
  wavenumber $k_x$ and spanwise wavenumber $k_z$. 
  (A) At fixed spanwise wavenumber $k_z=\{0.0,0.25,0.5,0.6\}$ top to bottom. 
  (B) At fixed streamwise wavenumber $k_x=\{0.35,0.25,0.15,0.11667,0.1,0.075\}$ top to bottom.}
\label{LinGrowthRates}
\end{figure}
%%%%%%%%%%%%%%%%%%%%%%%%%%%%%%%%%%%%%%%%%%%%%%%%%%%%%%%%%%%%%%%%%%%%%
%%\clearpage
\begin{figure}[ht]
 {\resizebox{0.45\textwidth}{!}{\includegraphics{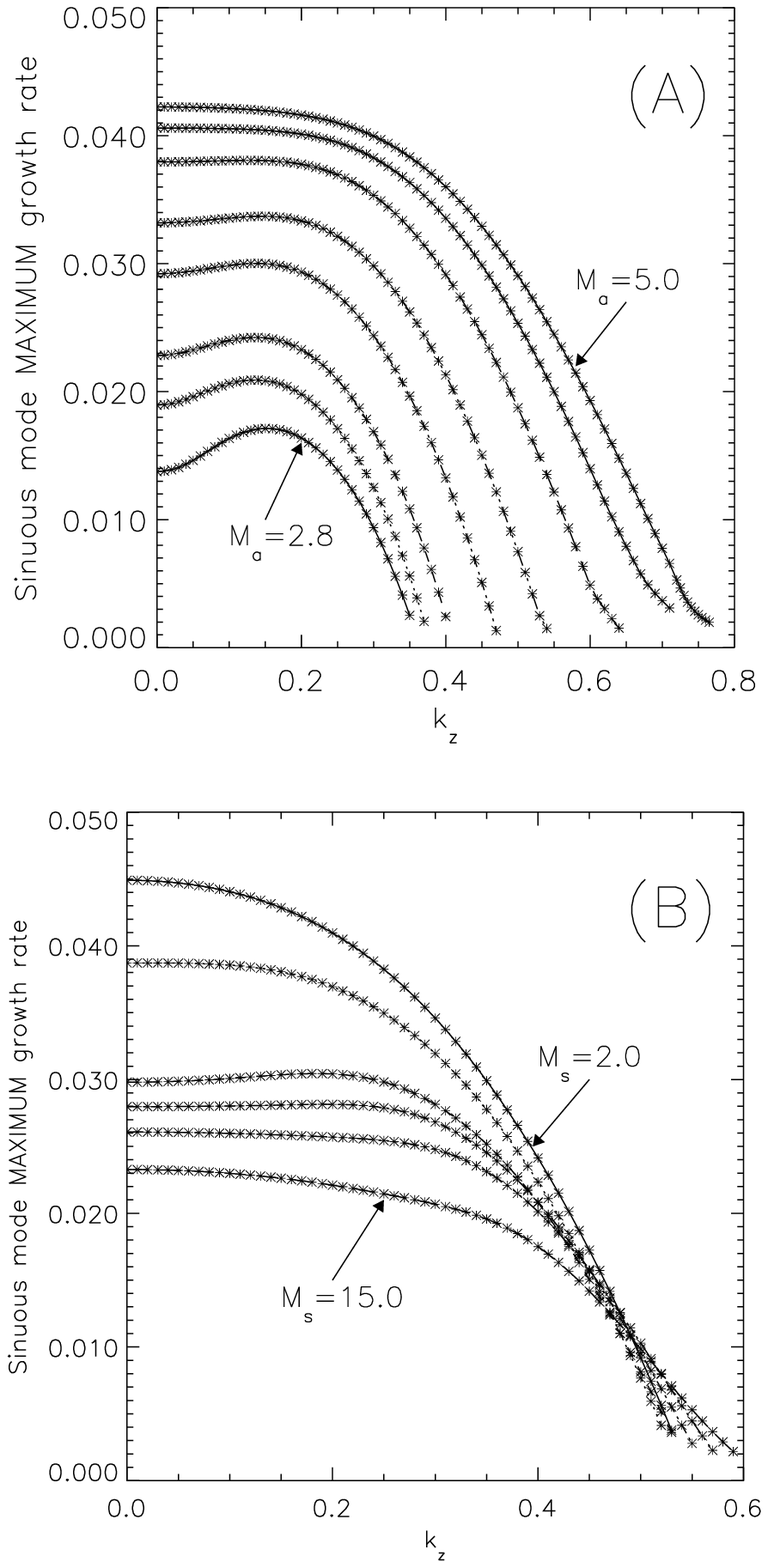}}}
  \caption{Maximum growth rates of the ideal sinuous mode in a
      wake -- current sheet system versus spanwise wavenumber for different 
      $M_s$ and $M_a$. Note that the streamwise wavenumber $k_x$ varies from curve to curve.
      (A) At fixed $M_s=3$, curves corresponding to
       $M_a=\{5.0, 4.5, 4.0, 3.5, 3.25, 3.0, 2.9, 2.8 \}$ top to bottom, 
       (B) At fixed $M_a=3.5$, curves corresponding to $M_s=\{2.0, 2.5, 3.5, 4.0, 5.0, 15.0 \}$ top to
       bottom.}
\label{GammaMax_Ms_Ma}
\end{figure}
%%%%%%%%%%%%%%%%%%%%%%%%%%%%%%%%%%%%%%%%%%%%%%%%%%%%%%%%%%%%%%%%%%%%%
\clearpage
\begin{figure}[ht]
%% {\resizebox{0.85\textwidth}{!}{\includegraphics{Fig6.eps}}}
\caption{2.5D MHD simulations of a wake-current sheet configuration
         with $M_s=3$, $M_a=5$, $\eta=10^{-4}$. 
         Plasma density (left panels) and cross-stream velocity (right panels)
         at times $T={100, 175,200,250,300,500}$. 
	 Two sinuous modes were excited and shocks of
         different strengths emerge at $T=300$.}
\label{Plate25D}
\end{figure}
%%%%%%%%%%%%%%%%%%%%%%%%%%%%%%%%%%%%%%%%%%%%%%%%%%%%%%%%%%%%%%%%%%%%%
\clearpage
\begin{figure}[ht]
 {\resizebox{0.99\textwidth}{!}{\includegraphics{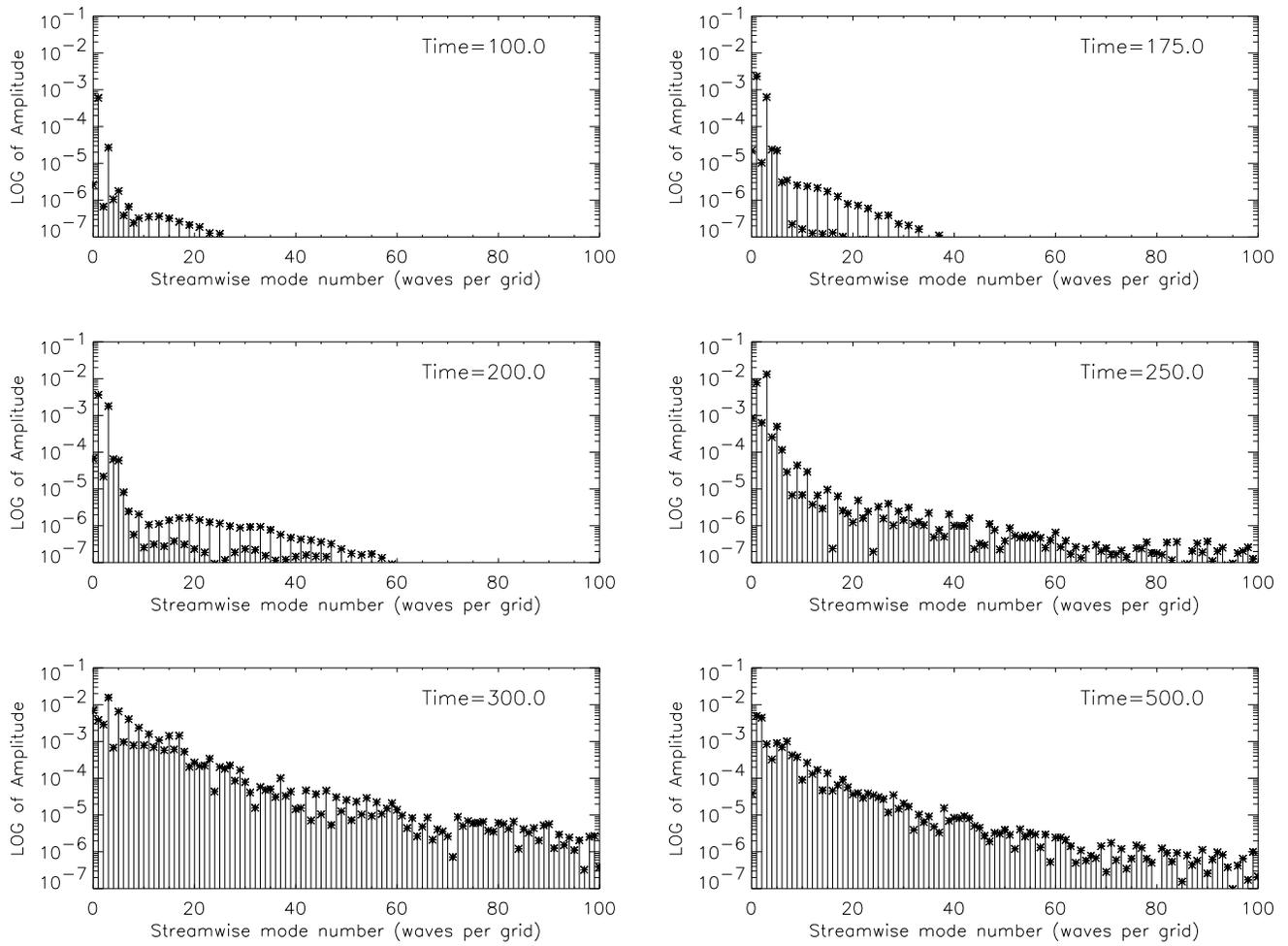}}}
\caption{Fourier spectrum of cross-stream velocity corresponding to the
2.5D run from Fig.~\ref{Plate25D}.}
\label{Spectra25D}
\end{figure}
%%%%%%%%%%%%%%%%%%%%%%%%%%%%%%%%%%%%%%%%%%%%%%%%%%%%%%%%%%%%%%%%%%%%%
\clearpage
\begin{figure}[ht]
%%
% {\resizebox{0.32\textwidth}{!}{\includegraphics{Fig8.eps}}}
 \caption{Vertical kinetic energy, total kinetic energy and magnetic energy versus time for the 2.5D run 
          from Fig.~\ref{Plate25D}.}
\label{Energies25D}
\end{figure}
%%%%%%%%%%%%%%%%%%%%%%%%%%%%%%%%%%%%%%%%%%%%%%%%%%%%%%%%%%%%%%%%%%%%%
\newpage
\begin{figure}[ht]
{\resizebox{0.49\textwidth}{!}{\includegraphics{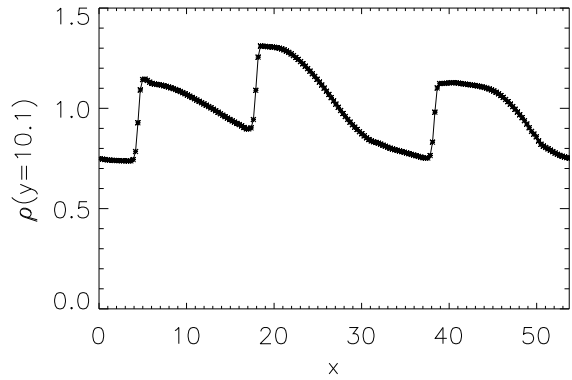}}}
 \caption{Plasma density distribution along the streamwise direction, far into
          the fast stream regions ($y=10.1$) at $T=300$, as obtained from 
          Fig.~\ref{Plate25D}. The fast shocks carry density jumps up to factors of two.}
\label{RHOCut25D}
\end{figure}
%%%%%%%%%%%%%%%%%%%%%%%%%%%%%%%%%%%%%%%%%%%%%%%%%%%%%%%%%%%%%%%%%%%%%
%%\newpage
\begin{figure}[ht]
{\resizebox{0.49\textwidth}{!}{\includegraphics{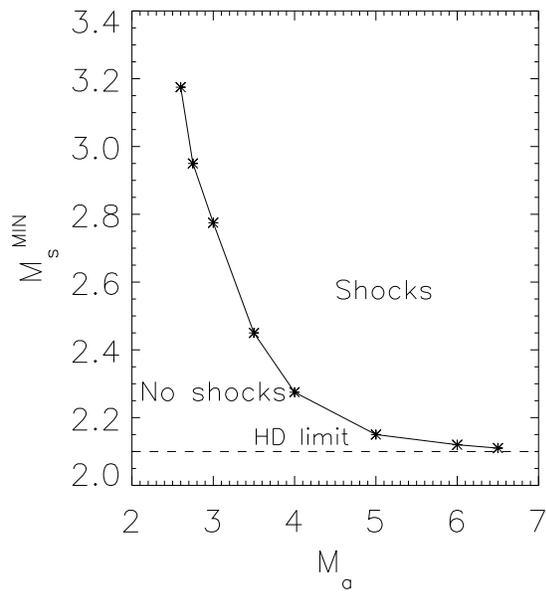}}}
\caption{Threshold value of sonic Mach number necessary for shock formation
         versus Alfv\'en Mach number.}
\label{ThresholdMs}
\end{figure}
%%%%%%%%%%%%%%%%%%%%%%%%%%%%%%%%%%%%%%%%%%%%%%%%%%%%%%%%%%%%%%%%%%%%%
\begin{figure}[ht]
%%%
%%% {\resizebox{0.99\textwidth}{!}{\includegraphics{Fig11.eps}}}

 \caption{3D MHD simulations of wake-current sheets. First column: $M_s=3$, 
          $M_a=5$, two modes excited: $(k_x=0.35, \,\,k_z=0)$ and $(k_x=0.35/3=0.11667, \,\, k_z=0.125)$.
          Second column: as in first column, only the 3D mode excited initially. Third
          column: $M_s=3$, $M_a=3.5$, initiated with its dominant three-dimensional 
          sinuous mode $(k_x=0.375, \,\,
          k_z=0.15)$. Density isosurfaces at consecutive times (top to bottom) are shown.}
\label{Plate3D}
\end{figure}
%%%%%%%%%%%%%%%%%%%%%%%%%%%%%%%%%%%%%%%%%%%%%%%%%%%%%%%%%%%%%%%%%%%%%
\begin{figure}[ht]
%%%
%%% {\resizebox{0.32\textwidth}{!}{\includegraphics{Fig12.eps}}}
%%%
\caption{Evolution of vertical kinetic energies for all three 3D runs in 
         Fig.~\ref{Plate3D}.}
\label{Energies3D}
\end{figure}

\end{document}